\documentstyle{article}
\title{
\begin{flushright}
{\bf\normalsize   COLO-HEP-310 }  \\
\end{flushright}
\vskip 10pt
\bf Smooth Random Surfaces \\
    from Tight Immersions?
}

\author{ {\it C.F. Baillie} \\
         Physics Dept. \\
         University of Colorado\\
         Boulder, CO 80309\\
         USA\\
         \\
         {\it D.A. Johnston} \\
         Dept. of Mathematics\\
         Heriot-Watt University\\
         Riccarton\\
         Edinburgh, EH14 4AS\\
         Scotland}
\begin{document}
  \maketitle
                      {\Large
                      \begin{abstract}
%
We investigate actions for dynamically triangulated random surfaces
that consist of a gaussian or area term plus the {\it modulus} of the gaussian
curvature and compare their behavior with both gaussian plus extrinsic
curvature and ``Steiner'' actions.
%
                        \end{abstract} }
%
  \thispagestyle{empty}
%
%
  \newpage
%
                  \pagenumbering{arabic}
Considerable effort has recently been devoted to exploring
modifications of the discretized Polyakov partition function \cite{1}
for a random surface
\begin{equation}
Z = \sum_{T} \int \prod_{i=1}^{N-1} d X_i^{\mu} \exp (- S_g),
\label{e2}
\end{equation}
where the sum over triangulations $\sum_{T}$ means that we have, in effect, a
fluid surface.
$S_g$ is just a simple gaussian action
\begin{equation}
S_g = {1 \over 2} \sum_{<ij>} (X_i^{\mu} - X_j^{\mu})^2,
\label{e3}
\end{equation}
where the $X$'s live at the vertices of the triangulation and the sum $<ij>$ is
over all the edges.
Earlier work \cite{2} had made it clear that this action (and variations,
such as area and edge length actions \cite{3,4})
failed to lead to a sensible continuum theory because the string
tension did not scale so, inspired by analytical work on QCD strings and
biological membranes \cite{5},
an extrinsic curvature or ``stiffness'' term was added
\begin{equation}
S_e = \sum_{\Delta_i, \Delta_j} ( 1 - n_i \cdot n_j),
\label{e4}
\end{equation}
where $n_i, n_j$ are the normals on neighboring triangles $\Delta_i, \Delta_j$.
Simulations of $S_g + \lambda S_e$, called the gaussian plus
extrinsic curvature (GPEC) action, seemed to indicate that there was a second
order
phase transition (the ``crumpling transition'')
from a small $\lambda$ crumpled phase to a large $\lambda$ smooth phase at
which one might hope to
define a non-trivial continuum theory \cite{6}. More recent simulations of
larger surfaces
suggest, however, that the transition is not second order \cite{7,8} \footnote{
Unlike the
GPEC action on {\it rigid} lattices, which almost certainly is \cite{9}.} but
that
the string tension may possibly still be scaling correctly.

Simpler spin systems on dynamical triangulations, such as Ising and Potts
models \cite{9a}, provide some reassurance that a non-trivial theory may be
lurking
at the crumpling transition because
they display third order transitions and still have a sensible continuum limit.
Nonetheless,
the rather murky behavior seen in \cite{7,8} prompts the question of whether
actions
with a sharper phase transition can be found. One geometrically appealing
suggestion
was the Steiner action put forward in \cite{10}, and simulated microcanonically
in \cite{11}
\begin{equation}
S_{steiner} = {1 \over 2} \sum_{<ij>} | X_i^{\mu} - X_j^{\mu} | \theta
(\alpha_{ij}),
\label{e4a}
\end{equation}
where
$\theta(\alpha_{ij}) = | \pi - \alpha_{ij} |$ and $\alpha_{ij}$ is the angle
between the
embedded neighboring triangles with common link $<ij>$.  It was pointed out in
\cite{12}
that the grand canonical partition function diverged for $S_{steiner}$ alone,
so we conducted some
exploratory simulations of various actions combining edge-length, area or
gaussian terms with
$S_{steiner}$ finding particularly sharp transitions for $Area + \lambda
S_{steiner}$ and $S_g + \lambda S_{steiner}$ actions \cite{13}.
As this initial work is on the same small triangulations that indicated a
second order transition
for the GPEC action we cannot claim this is strong evidence for a sharp
transition with the Steiner actions,
without further results from larger lattices \cite{14}.

In this paper we will explore a further possibility for a random surface action
which, like the Steiner term, is
a natural object to consider from a geometrical point of view. For a curve $C$
embedded in three dimensions it was shown by Fenchel that
\begin{equation}
{1 \over \pi} \int_C | \kappa | ds \ge 2
\label{e5}
\end{equation}
where $\kappa$ is the curvature and the equality holds when $C$ is a plane
convex curve \cite{15}.
For a surface $M$ imbedded in three dimensions the Gauss-Bonnet theorem tells
us that
\begin{equation}
{1 \over 2 \pi} \int_M K dS =  \chi (M)
\label{e6}
\end{equation}
where $K$ is now the Gaussian curvature of the surface and $\chi$ is the Euler
characteristic. As this is
a topological invariant it tells us nothing about the configuration of the
surface. To get the equivalent
of Fenchel's  equ.(\ref{e5}) we take a modulus sign in equ.(\ref{e6}) and find
\footnote{As a particular case of the theorem: Let $M^n$ be a compact oriented
$C^{\infty}$ manifold
immersed in $E^{n+N}$, such that the total absolute curvature equals 2. Then
$M^n$ belongs to
a linear subvariety of dimension $n+1$, and is imbedded as a convex
hypersurface in $E^{n+1}$. The converse
is also true \cite{15}.}
\begin{equation}
{1 \over 2 \pi} \int_M | K | dS \ge 4 -  \chi (M)
\label{e7}
\end{equation}
with the equality holding when the surface is imbedded as a convex surface in
three dimensional
space.
This term discretizes to
\begin{equation}
S_{tight} = \sum_i | 2 \pi   - \sum_{j(i)}  \phi_{ij}  |
\label{e8}
\end{equation}
where the outer sum is over all the vertices of the triangulation and the inner
sum is round the neighbors
$j$ of a node $i$. $\phi_{ij}$ is the angle subtended by the $j$th triangle at
the $i$th vertex -- see Fig.1.
The behavior of $S_{tight}$ was measured for the GPEC action in \cite{8}
(rather than being included in the action),
where it was found to correlate closely with the extrinsic curvature - dropping
sharply in value at the
crumpling transition. This raises the hope that including $S_{tight}$ in the
dynamics may be sufficient to
give a crumpling transition without the assistance of an extrinsic curvature
term.

We will examine the phase structure of both $S_1 = S_g + \lambda S_{tight}$
\begin{equation}
S_1 = {1 \over 2} \sum_{<ij>} (X_i^{\mu} - X_j^{\mu})^2 + \lambda \sum_i | 2
\pi   - \sum_{j(i)}  \phi_{ij}  |
\label{e9}
\end{equation}
and $S_2 = Area + \lambda S_{tight}$
\begin{equation}
S_2 = \sum_{\Delta} A_{\Delta} + \lambda \sum_i | 2 \pi   - \sum_{j(i)}
\phi_{ij}  |
\label{e9a}
\end{equation}
where $A_{\Delta}$ is the area of a triangle $\Delta$, in what follows.
To do this we employ our by now standard set of observables.
We included a local factor in
the measure for
compatibility with our earlier simulations which can be exponentiated to give
\begin{equation}
S_m = { d \over 2 } \sum_i \log ( q_i ),
\label{e10}
\end{equation}
where $q_i$ is the number of neighbors of point $i$, and $d=3$ dimensions.
We thus simulated $S_{1,2} + S_m$.
We measured $<S_m>$ and the mean maximum number of neighbors $<max(q_i)>$ to
get some idea
of the behavior of the intrinsic geometry.
The extrinsic geometry
was observed by measuring $<S_{tight}>$ and its
associated specific heat
\begin{equation}
C = {\lambda^2 \over N} \left( < S_{tight}^2 > - < S_{tight} >^2
\right)\label{e11}
\end{equation}
as well as the gyration radius $X2$, a measure of the mean size of the
surface as seen in the embedding space,
\begin{equation}
X2 = { 1 \over 9 N (N -1)} \sum_{ij} \left( X^\mu_i - X^\mu_j \right)^2
q_i q_j.
\label{e12}
\end{equation}
For $S_1$ the expectation value of $S_g$ can be shown to be $d( N-1)/ 2$
by exactly the same argument that is used to give this result for the GPEC
action, because
$S_{tight}$ shares the scale invariance of the extrinsic curvature term in the
GPEC action.
This serves as a useful check of equilibration in this case. A further useful
check is
provided by removing the modulus sign in equ.(\ref{e8}) which should then sum
to give $4 \pi$ for
every configuration by the Gauss-Bonnet theorem.

The simulation used a Monte Carlo procedure
which we have described in some detail elsewhere \cite{16}. It first goes
through the mesh moving the $X$'s, carrying out a Metropolis
accept/reject at each step, and then goes through the mesh again
carrying out the ``flip'' moves on the links, again applying a
Metropolis accept/reject at each stage. The entire procedure
constitutes a sweep. Due to the correlated nature of the
data, a measurement was taken every tenth sweep and binning
techniques were used to analyze the errors. We carried out 10K
thermalization sweeps  followed by 30K  measurement sweeps for
each data point. The acceptance for the $X$ move
was monitored and the size of the shift was adjusted to maintain an
acceptance of around 50 percent. The acceptance for the flip move was
also measured, but in this case there is nothing to adjust, so as for
GPEC actions this dropped with increasing $\lambda$ (but
was still appreciable even for quite large $\lambda$).

If we look at the numerical results for $S_1$ ($S_g + \lambda S_{tight})$ in
Table 1 first
\vskip 10pt
\hoffset=0truein
\voffset=0truein
\centerline{
\hbox{\vbox{ \tabskip=0pt \offinterlineskip
\def\tablerule{\noalign{\hrule}}
\halign{\strut#& \vrule#&
\hfill#\hfill &\vrule#& \hfill#\hfill &\vrule#&
\hfill#\hfill &\vrule#& \hfill#\hfill &\vrule#&
\hfill#\hfill &\vrule#& \hfill#\hfill &\vrule#&
\hfill#\hfill &\vrule#& \hfill#\hfill &\vrule#
\tabskip=0pt\cr
\tablerule
\omit&height2pt& \omit&height2pt&
\omit&height2pt& \omit&height2pt&
\omit&height2pt& \omit&height2pt&
\omit&height2pt& \omit&height2pt&
\omit&\cr
&&\enskip $\lambda$ \enskip
&&\enskip sweeps \enskip
&&\enskip $S_g$ \enskip
&&\enskip $S_m$ \enskip
&&\enskip $S_{tight}$ \enskip
&&\enskip $C_{tight}$ \enskip
&&\enskip $X2$ \enskip
&&\enskip $max(q_i)$ \enskip
&\cr
\omit&height2pt& \omit&height2pt&
\omit&height2pt& \omit&height2pt&
\omit&height2pt& \omit&height2pt&
\omit&height2pt& \omit&height2pt&
\omit&\cr
\tablerule
\omit&height2pt& \omit&height2pt&
\omit&height2pt& \omit&height2pt&
\omit&height2pt& \omit&height2pt&
\omit&height2pt& \omit&height2pt&
\omit&\cr
&& 0.500 && 30K &&   106.62(0.03) &&   122.79(0.00) &&   109.94(0.03) &&
0.46(  0.00) &&     2.40(0.01) &&    12.07(0.00) &\cr
&& 1.000 && 30K &&   106.33(0.01) &&   123.60(0.00) &&    66.40(0.00) &&
0.77(  0.00) &&     2.54(0.01) &&    10.92(0.00) &\cr
&& 1.250 && 30K &&   106.34(0.05) &&   123.76(0.00) &&    54.96(0.01) &&
0.84(  0.00) &&     2.53(0.01) &&    10.70(0.00) &\cr
&& 1.500 && 30K &&   106.45(0.03) &&   123.83(0.00) &&    46.70(0.02) &&
0.86(  0.00) &&     2.80(0.04) &&    10.59(0.00) &\cr
&& 1.750 && 30K &&   106.58(0.10) &&   123.89(0.00) &&    40.77(0.02) &&
0.91(  0.00) &&     2.77(0.05) &&    10.52(0.00) &\cr
&& 2.000 && 30K &&   107.25(0.21) &&   123.91(0.00) &&    36.10(0.05) &&
0.93(  0.00) &&     2.94(0.12) &&    10.49(0.00) &\cr
&& 2.250 && 30K &&   106.28(0.02) &&   123.92(0.00) &&    32.11(0.01) &&
0.90(  0.00) &&     2.16(0.00) &&    10.48(0.00) &\cr
&& 2.500 && 30K &&   106.34(0.07) &&   123.91(0.00) &&    29.42(0.02) &&
0.90(  0.00) &&     2.17(0.01) &&    10.48(0.00) &\cr
&& 3.000 && 30K &&   106.52(0.08) &&   123.90(0.00) &&    25.18(0.00) &&
0.82(  0.00) &&     2.14(0.01) &&    10.49(0.00) &\cr
&& 3.500 && 30K &&   106.48(0.10) &&   123.88(0.00) &&    22.53(0.01) &&
0.78(  0.00) &&     2.15(0.01) &&    10.51(0.00) &\cr
&& 4.000 && 30K &&   106.02(0.05) &&   123.84(0.00) &&    20.56(0.01) &&
0.73(  0.00) &&     2.12(0.02) &&    10.53(0.00) &\cr
&& 4.500 && 30K &&   106.08(0.37) &&   123.80(0.00) &&    19.15(0.00) &&
0.68(  0.00) &&     2.20(0.07) &&    10.57(0.01) &\cr
\omit&height2pt& \omit&height2pt&
\omit&height2pt& \omit&height2pt&
\omit&height2pt& \omit&height2pt&
\omit&height2pt& \omit&height2pt&
\omit&\cr
\tablerule
}}}}
\smallskip
\centerline{Table 1}
\centerline{Results for $S_1, N=72$}

\vskip 5pt
\noindent
we can see that $S_{tight}$ does indeed drop off with increasing $\lambda$ just
like the extrinsic
curvature. The behavior of $S_m$ and $max(q_i)$ is also reminiscent of the GPEC
action,
with the internal geometry becoming more regular with increasing $\lambda$.
However, the specific
heat $C$ shows only a modest cusp at around $\lambda=2.00$ as can be seen in
Fig.2, which should be
contrasted with the larger peaks seen on these small meshes for both the GPEC
and Steiner actions.
Similarly $X2$, plotted in Fig.3, shows no sign of a crumpling transition, with
only a small
increase in the region of the cusp in $C$, before it rapidly drops off. The
value of the gaussian
term is close to the expected $d(N-1)/2$, assuring us that the results are
equilibrated and both the metropolis and flip acceptances are reasonable for
all the
values of $\lambda$ simulated, so we can be sure that the simulation is
performing as it should.
We also measured the value of the gaussian curvature using $S_{tight}$ with the
modulus sign removed
and found, as expected (our surfaces have spherical topology), $4 \pi$ for
every surface generated.
Visual inspection of ``snapshots'' of the surfaces that arise in the simulation
confirms the
absence of a phase transition, with surfaces looking similar for all of the
$\lambda$ values simulated.
One of these for $\lambda = 4.0$, but which is typical of all the others, is
shown in Fig.4 and is obviously
not smooth. Even at the largest $\lambda$ values simulated the surfaces are
still some way from
satisfying the lower bound of $4 \pi$ on $S_{tight}$, so it would appear that
the disordering effect of
$S_g$ overcomes $S_{tight}$ for all $\lambda$.

The behavior of $S_2$ ($Area + \lambda S_{tight})$ is rather bizarre, as can be
seen in Table 2.
\vskip 10pt
\hoffset=0truein
\voffset=0truein
\centerline{
\hbox{\vbox{ \tabskip=0pt \offinterlineskip
\def\tablerule{\noalign{\hrule}}
\halign{\strut#& \vrule#&
\hfill#\hfill &\vrule#& \hfill#\hfill &\vrule#&
\hfill#\hfill &\vrule#& \hfill#\hfill &\vrule#&
\hfill#\hfill &\vrule#& \hfill#\hfill &\vrule#&
\hfill#\hfill &\vrule#& \hfill#\hfill &\vrule#
\tabskip=0pt\cr
\tablerule
\omit&height2pt& \omit&height2pt&
\omit&height2pt& \omit&height2pt&
\omit&height2pt& \omit&height2pt&
\omit&height2pt& \omit&height2pt&
\omit&\cr
&&\enskip $\lambda$ \enskip
&&\enskip sweeps \enskip
&&\enskip $Area$ \enskip
&&\enskip $S_m$ \enskip
&&\enskip $S_{tight}$ \enskip
&&\enskip $C_{tight}$ \enskip
&&\enskip $X2$ \enskip
&&\enskip $max(q_i)$ \enskip
&\cr
\omit&height2pt& \omit&height2pt&
\omit&height2pt& \omit&height2pt&
\omit&height2pt& \omit&height2pt&
\omit&height2pt& \omit&height2pt&
\omit&\cr
\tablerule
\omit&height2pt& \omit&height2pt&
\omit&height2pt& \omit&height2pt&
\omit&height2pt& \omit&height2pt&
\omit&height2pt& \omit&height2pt&
\omit&\cr
&& 0.500 && 30K &&   106.05(0.03) &&   122.26(0.00) &&   119.51(0.09) &&
0.70(  0.01) &&    10.67(0.17) &&    12.66(0.00) &\cr
&& 1.000 && 30K &&   105.81(0.13) &&   122.88(0.00) &&    61.28(0.43) &&
0.86(  0.01) &&    22.60(3.46) &&    11.91(0.01) &\cr
&& 1.500 && 30K &&   105.78(0.18) &&   122.95(0.01) &&    38.55(0.28) &&
0.63(  0.01) &&   113.70(11.84) &&    11.80(0.02) &\cr
&& 1.750 && 30K &&   106.11(0.38) &&   122.93(0.02) &&    34.01(0.50) &&
0.69(  0.01) &&   141.42(42.03) &&    11.89(0.05) &\cr
&& 2.000 && 30K &&   105.50(0.10) &&   122.73(0.03) &&    31.63(0.14) &&
0.79(  0.03) &&    33.69(10.03) &&    12.42(0.07) &\cr
&& 2.250 && 30K &&   104.85(0.08) &&   122.70(0.01) &&    29.55(0.06) &&
0.87(  0.03) &&    25.39(3.94) &&    12.57(0.03) &\cr
&& 3.000 && 30K &&   106.04(0.06) &&   122.34(0.00) &&    22.21(0.02) &&
0.61(  0.00) &&    10.31(0.10) &&    13.60(0.00) &\cr
&& 3.500 && 30K &&   105.59(0.09) &&   122.19(0.00) &&    20.10(0.01) &&
0.56(  0.00) &&    10.28(0.03) &&    14.08(0.01) &\cr
&& 4.000 && 30K &&   106.68(0.25) &&   122.12(0.01) &&    18.80(0.01) &&
0.54(  0.00) &&    10.50(0.15) &&    14.16(0.02) &\cr
&& 4.500 && 30K &&   106.26(0.23) &&   121.98(0.01) &&    17.68(0.01) &&
0.50(  0.00) &&    11.17(0.18) &&    14.67(0.02) &\cr
\omit&height2pt& \omit&height2pt&
\omit&height2pt& \omit&height2pt&
\omit&height2pt& \omit&height2pt&
\omit&height2pt& \omit&height2pt&
\omit&\cr
\tablerule
}}}}
\smallskip
\centerline{Table 2}
\centerline{Results for $S_2, N=72$}

\vskip 5pt
\noindent
Again $S_{tight}$ decreases with increasing $\lambda$ but there is no obvious
peak in the specific heat.
$max(q_i)$ now {\it increases} at large $\lambda$ and there is a huge peak in
$X2$ at around $\lambda=2.5$.
For $\lambda > 2.5$ the surfaces generated look rather similar to those
produced by $S_1$, but
near $\lambda=2.5$ they are very long, jointed linear structures such as that
in Fig.5.
On its own the area action gives surfaces that are collections of long thin
spikes
emanating from a central point, and this is also seen for $S_2$ at small
$\lambda$. It appears
that adding $S_{tight}$ gives (approximately) very long, thin ellipsoids which,
while
satisfying the convexity property, are not the generic smooth surfaces that we
envisaged.
{}From the evidence of the specific heat there is little sign of a phase
transition in this region
that might be used to define a continuum limit, though it is always possible
that a higher
order transition may be present.

Our conclusions are rather disappointing from the point of view of finding
candidate
random surface actions which might be used similarly to the GPEC action to hunt
for a non-trivial
continuum string theory. $S_1$ shows little sign of a transition at all, apart
from a modest
bump in the specific heat, and $S_2$, whilst adding to the bestiary of amusing
pathologies that have
been observed with various random surface models, also looks unpromising. It
may of course be
possible to incorporate $S_{tight}$ as an additional term in the GPEC action in
order to tune the couplings
to see if the approach to the continuum limit could be optimized. A further
possibility that might be
worth pursuing is looking at the effect of self-avoidance, which completely
changes the behavior of GPEC actions \cite{17}.

This work was supported in part by NATO collaborative research grant CRG910091.
CFB is supported by DOE under contract DE-FG02-91ER40672
and by NSF Grand Challenge Applications Group Grant ASC-9217394.
The computations were performed on workstations at
Heriot-Watt University.
We would like to thank
R.D. Williams for help in developing initial versions of the dynamical
mesh code.

\vfill
\eject

\vfill
\eject

\centerline{\bf Figure Captions}
\begin{description}
\item[Fig. 1.]
The neighbors and angles used in defining $S_{tight}$.
\item[Fig. 2.]
The specific heat $C$ for action $S_1$.
\item[Fig. 3.]
The gyration radius $X2$ for action $S_1$.
\item[Fig. 4.]
A snapshot of a mesh generated by $S_1$ with $\lambda=4.5$.
\item[Fig. 5.]
A snapshot of a mesh generated by $S_2$ with $\lambda=2.5$.
\end{description}

\end{document}